\documentclass[doublecol]{epl2}

\title{Evolution of public cooperation on interdependent networks:\\The impact of biased utility functions}
\shorttitle{Evolution of public cooperation on interdependent networks: The impact of biased utility functions}

\author{Zhen Wang,\inst{1,2} Attila Szolnoki,\inst{3} Matja{\v z} Perc\inst{4}}
\shortauthor{Wang et al.}

\institute{\inst{1}School of Physics, Nankai University, Tianjin 300071, China\\
\inst{2}Department of Physics, Hong Kong Baptist University, Kowloon Tong, Hong Kong\\
\inst{3}Research Institute for Technical Physics and Materials Science, P.O. Box 49, H-1525 Budapest, Hungary\\
\inst{4}Faculty of Natural Sciences and Mathematics, University of Maribor, Koro{\v s}ka  cesta 160, SI-2000 Maribor, Slovenia}

\pacs{87.23.Ge}{Dynamics of social systems}
\pacs{87.23.Kg}{Dynamics of evolution}
\pacs{89.75.Fb}{Structures and organization in complex systems}

\abstract{We study the evolution of public cooperation on two interdependent networks that are connected by means of a utility function, which determines to what extent payoffs in one network influence the success of players in the other network. We find that the stronger the bias in the utility function, the higher the level of public cooperation. Yet the benefits of enhanced public cooperation on the two networks are just as biased as the utility functions themselves. While cooperation may thrive on one network, the other may still be plagued by defectors. Nevertheless, the aggregate level of cooperation on both networks is higher than the one attainable on an isolated network. This positive effect of biased utility functions is due to the suppressed feedback of individual success, which leads to a spontaneous separation of characteristic time scales of the evolutionary process on the two interdependent networks. As a result, cooperation is promoted because the aggressive invasion of defectors is more sensitive to the slowing down than the build-up of collective efforts in sizable groups.}

\begin{document}

\maketitle

\section{Introduction}

The study of evolutionary games on networks and graphs (see \cite{szabo_pr07} for a comprehensive review) has proven very gratifying in terms of improving our understanding of the emergence and sustenance of cooperation among selfish and unrelated individuals. Following the seminal discovery that spatial structure may, unlike well-mixed populations, maintain cooperation even in the most challenging prisoner's dilemma game \cite{nowak_n92b}, and the many groundbreaking discoveries concerning the statistical mechanics of complex networks and the dynamical processes taking place on them \cite{albert_rmp02, boccaletti_pr06}, the study of evolutionary games on small-world \cite{kim_bj_pre02}, scale-free \cite{santos_prl05}, coevolving \cite{ebel_pre02, zimmermann_pre04} and hierarchical \cite{vukov_pre05} networks, to name but a few, now appears as having been the logical next step. From these and many other related studies, we have learnt that scale-free networks might be the missing link to cooperation by virtually all main social dilemmas \cite{santos_pnas06}, and that this is a very robust evolutionary outcome \cite{poncela_njp07}, although  not immune to the normalization of payoffs \cite{santos_jeb06, masuda_prsb07, tomassini_ijmpc07, szolnoki_pa08} and targeted removal of nodes \cite{perc_njp09}. Moreover, heterogeneity in general, \textit{i.e.} not just in terms of players having different degree within a network, proved to be very effective in maintaining high levels of cooperation in the population \cite{szolnoki_epl07, ren_pre07, guan_pre07, perc_pre08, santos_n08, gomez-gardenes_epl11, santos_jtb12}, and indeed many coevolutionary rules have been introduced that may generate such states spontaneously \cite{pacheco_prl06, szolnoki_epl08, fu_pre08, zhang_j_pa11, poncela_ploso08, wu_t_pre09, wu_t_epl09, poncela_njp09,zhang_j_pa10,dai_ql_njp10,lin_yt_pa11} (see \cite{perc_bs10} for a review).

Recently, however, it has been emphasized that, although research on complex networks has been flourishing and has become an integral part of many branches of physics, the focus is predominantly still on single (or isolated) networks \cite{buldyrev_n10}. In many ways this approach can be considered as rather limited, since real networks are simultaneously present and influence each other, and should thus be treated as interdependent networks. Several examples attesting to this fact are given in \cite{parshani_prl10}, while specifically for evolutionary games, it is possible to argue that the interaction network of players may be just as important for their success as the network of institutions providing the funding, or the network of governmental bodies overseeing that everybody is obeying the rules. More generally, the success of a player in a given network may not depend just on the players in that same network, but may also depend on a player that is a member of another network, thus lending ample justification to studying the outcome of evolutionary games on interdependent networks. Until now, however, seminal works on interdependent networks have shown that seemingly irrelevant changes in one network can have catastrophic and very much unexpected consequence in another network \cite{buldyrev_n10}.

In this paper, we study the evolution of public cooperation on two interdependent networks, which are linked together by means of a utility function that is defined as a combination of individual payoffs of related pairs of players selected from different graphs. In general, payoffs represent utilities that players try to maximize by adopting strategies from others. There are, however, several realistic situations when our actions are not motivated solely by our own wellbeing, but may also depend on the impact they will have on others, \textit{e.g.} the family or the closest collaborators. The determination of an accurate utility function is therefore demanding, typically involving the consideration of fraternity, other-regarding preferences, or simply the behavior of relatives in biological systems \cite{taylor_c_tpb06, taylor_c_e07, xianyu_b_pa10, szabo_jtb11}. Here, conceptually differently, we use the concept of utility to link together two networks, for convenience denoted as networks A and B, which therefore become interdependent in a way that is paramount for the outcome of the game. In particular, we define the utility of each player as a biased sum of the payoff of the player itself and the payoff of the corresponding player in the other network. In this way, players in network A consider the payoffs of players in network B to be more relevant than their own, while players in network B consider their own payoffs more prominently than those of the players in network A. In order to focus explicitly on the impact of this interdependence, and to avoid potential effects stemming from complex networks, we use the square lattice topology for both networks A and B. This also enables us to compare the obtained results accurately with those reported previously on a single network \cite{szolnoki_pre09c}. Moreover, due to the exactly defined locations of all the players, the potential complications with defining who are the corresponding players in the two networks are naturally alleviated. Interestingly, we find that the interdependence by means of biased utility promotes the evolution of cooperation, yet that the extent of this promotion itself is heavily biased in the two networks. Accurate results that will be presented below firmly attest to the fact that the integration of interdependent networks and evolutionary games offers new ways of understanding the successful evolution of cooperation, as well as provides ample opportunities for further research along this line.

The remainder of this letter is organized as follows. First, we describe the considered  public goods game and the interdependence of the two networks due to the biased definition of the utility function. Next we present the main results, whereas lastly we summarize them and discuss their implications.

\section{Model definition}

The public goods game on both networks is staged on a $L \times L$ square lattice with periodic boundary conditions, where players are arranged into overlapping groups of size $G=5$. Every player is thus surrounded by its $k=G-1$ nearest neighbors and is a member in $g=G$ different groups. Initially each player on site $x$ in network A and on site $x^\prime$ in network B is designated either as a cooperator or defector with equal probability. The accumulation of payoffs $P_x$ and $P_{x^\prime}$ on both networks follows the same standard procedure. Namely, in each group cooperators contribute $1$ to the public good while defectors contribute nothing. The sum of contributions is subsequently multiplied by the factor $r>1$, reflecting the synergetic effects of cooperation, and the resulting amount is equally shared amongst the $G$ group members. In each group the payoff obtained is $P_x^g$ on network A and $P_{x^\prime}^g$ on network B, while the total amount received in all the groups is thus $P_x = \sum_g P_x^g$ and $P_{x^\prime} = \sum_g P_{x^\prime}^g$.

\begin{figure}
\centerline{\scalebox{0.29}[0.29]{\includegraphics{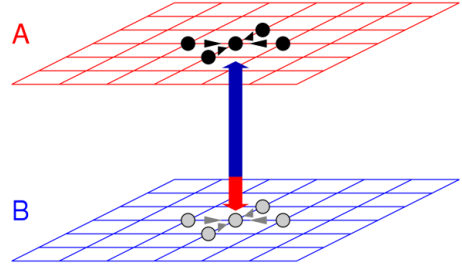}}}
\caption{Schematic presentation of the model. Players are arranged on two physically separated square lattices. Interdependence is introduced via the utility function, which determines the probability of strategy invasion within a lattice, and is calculated based not only on the player's own payoff but also on the payoff of its corresponding player in the other network. The two payoffs are considered in a biased manner, as marked by the different lengths of the vertical arrows. According to the scheme, the utility function in the upper network A (red) is determined predominantly by the payoffs in the lower network B (blue), while the utility function in the lower network B is only slightly influenced by the payoffs in the upper network A. Importantly, strategy invasions are possible from nearest neighbors only, as marked by the small arrows on both grids.}
\label{scheme}
\end{figure}

While the two networks are not physically connected, interdependence is introduced via the utility functions
\begin{equation}
U_x = \alpha P_x + (1-\alpha) P_{x^\prime}\,,\, U_{x^\prime} = (1-\alpha) P_{x^\prime} + \alpha P_x\,,
\label{utility}
\end{equation}
where $\alpha$ determines the bias in the consideration of payoffs collected by the corresponding players $x$ and $x^\prime$ in the two networks, as schematically depicted in Fig.~\ref{scheme}. At low $\alpha$ values player $x$ is guided predominantly by the payoff of player $x^\prime$, while at $\alpha=0.5$ both $P_x$ and $P_{x^\prime}$ are taken into consideration equally by both players $x$ and $x^\prime$. Evidently, for $\alpha > 0.5$ the roles are exchanged and the treatment becomes fully symmetric. It is also worth emphasizing that at $\alpha=1$ ($\alpha=0$) the game on network A (B) behaves identically as if played on a single network \cite{szolnoki_pre09c}, while the game on network B (A) is completely guided by the payoffs of players in network A (B).

Following the determination of utilities according to Eq.~\ref{utility}, strategy invasions are attempted between nearest neighbors on a given network (see Fig.~\ref{scheme}). Accordingly, on network A player $x$ can adopt the strategy $s_{y}$ of one of its randomly chosen nearest neighbors $y$ with a probability determined by the Fermi function
\begin{equation}
W(s_{y} \rightarrow s_{x})=\frac{1}{1+\exp[(U_x-U_y)/K]} \,,
\label{fermi}
\end{equation}
where the utility $U_{y}$ of player $y$ is evaluated identically as for player $x$. The probability of strategy invasion from player $y^\prime$ to player $x^\prime$ on network B is determined likewise, only that utilities $U_x^\prime$ and $U_y^\prime$ are used. Without loss of generality in Eq.~\ref{fermi} we set $K=0.5$ \cite{szolnoki_pre09c}, implying that players with a higher utility spread, but it is not impossible to adopt the strategy of a player having a lower utility. Simulations of the model were performed by means of a random sequential update, where each player on both networks had a chance to pass its strategy once on average during a Monte Carlo step (MCS). The linear system size was varied from $L=200$ to $800$ in order to avoid finite size effects, and the equilibration required up to $10^6$ MCS.

\section{Results}

\begin{figure}
\centerline{\scalebox{0.148}[0.148]{\includegraphics{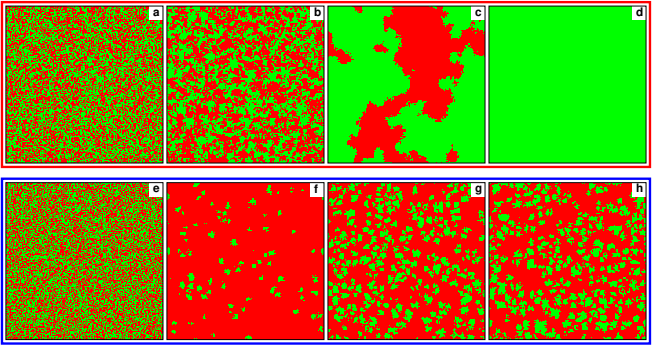}}}
\caption{Snapshots of the distribution of cooperators (green) and defectors (red) on the two interdependent square lattices at 0, 20, 2000 and 20000 MCS from left to right. Panels \textbf{a}-\textbf{d} (surrounded red) correspond to results obtained on network A, while panels \textbf{e}-\textbf{h} (surrounded blue) correspond to results obtained on network B. Parameter values are: $\alpha=0.01$, $r/G=0.76$ and $L=200$.}
\label{snaps}
\end{figure}

We start by presenting characteristic snapshots of the distribution of cooperators and defectors on the two networks in Fig.~\ref{snaps}. In order to demonstrate the impact of biased utility as effectively as possible, we use $\alpha=0.01$ and $r/G=0.76$. According to Eq.~\ref{utility}, this implies that the evolution on network A is guided predominantly (99\%) by the payoffs of players in network B, while the evolution on network B should be almost identical with the evolution on a single (isolated) square lattice. By focusing first on the snapshots in the bottom row of Fig.~\ref{snaps} (panels e-h), corresponding to the evolution on network B, it can indeed be observed that the outcome is very much similar to the one on an isolated lattice. In the stationary state (panels g and h) defectors dominate, while a relatively small fraction of cooperators is able to survive by forming compact clusters. This is in agreement with previous results obtained for a single square lattice, where cooperators can survive only if $r/G=r1 \geq 0.745$ \cite{szolnoki_pre09c}. Much more surprising, however, is the outcome in the upper row of Fig.~\ref{snaps} (panels a-d), corresponding to the evolution on network A. There the stationary state (panel d) is reached a full order of magnitude slower, yet instead of widespread defection, cooperators dominate completely. Thus, a strong bias in the utility function towards payoffs of players in the other network (network B in this case) significantly promotes the evolution of cooperation.

\begin{figure}
\centerline{\scalebox{0.5}[0.5]{\includegraphics{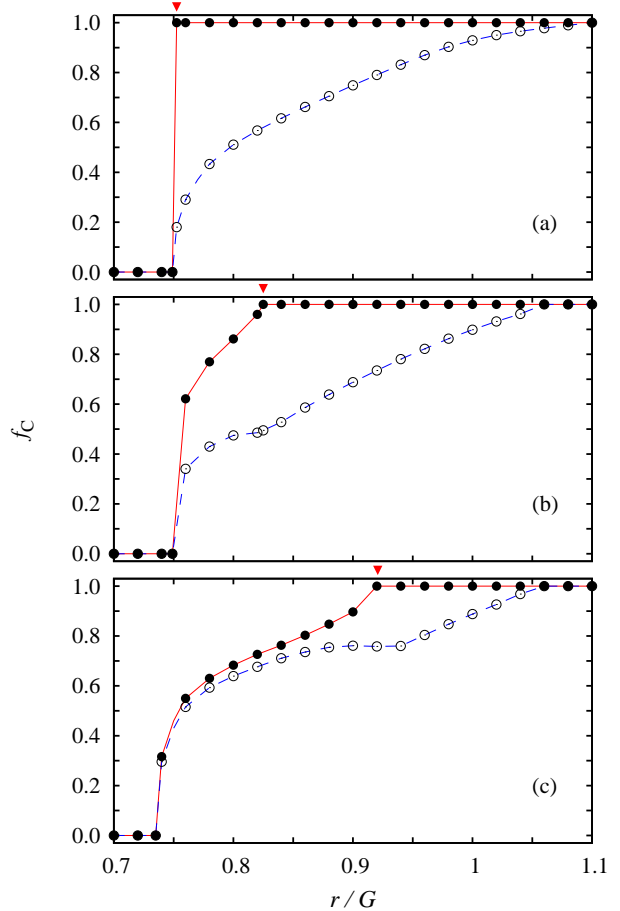}}}
\caption{Density of cooperators $f\rm{_C}$ in dependence on the normalized synergy factor $r/G$ as obtained on networks A (closed circles connected with a solid red line) and B (open circles connected with a dashed blue line) for $\alpha=0.01$ (a), 0.40 (b) and 0.49 (c). The critical values of the synergy factor $r2_c$, where the pure C phase is reached on network A, are marked by small red arrows at the top axis of each layer.}
\label{alpha}
\end{figure}

Results presented in Fig.~\ref{alpha} evidence clearly that the difference in the evolution of public cooperation on the two interdependent networks, as depicted by the snapshots in Fig.~\ref{snaps}, depends significantly on the value of $\alpha$. For $\alpha=0.01$ (panel a), where the bias in the utility function is the strongest, the difference is the largest, while for $\alpha=0.4$ (panel b), and even more so for $\alpha=0.49$ (panel c), the difference in the density of cooperators on the two networks is vanishing. According to the definition of utility (see Eq.~\ref{utility}), at $\alpha=0.5$ the difference vanishes completely (not shown). By trying to infer the aggregate level of cooperation on both networks, however, it can be deduced that the interval of $r$ where cooperators and defectors coexist is virtually independent of $\alpha$. With the aim of quantifying more accurately the impact of different $\alpha$ values on the evolution of public cooperation, we therefore focus on the normalized (with $G$) critical value of the synergy factor $r2_c$, where the pure C phase ($f\rm{_C}=1$) is reached on network A. These critical values are marked by small red arrows in Fig.~\ref{alpha}.

\begin{figure}
\centerline{\scalebox{0.5}[0.5]{\includegraphics{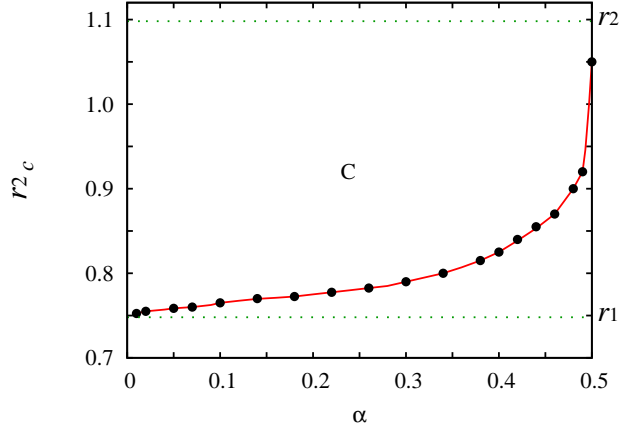}}}
\caption{Normalized critical synergy factor $r2_c$, required for full cooperator dominance on network A, in dependence on $\alpha$. For comparison, dotted green lines depict the normalized value of the synergy factor $r1=0.745$ ($r2=1.1$) where the full D (full C) is reached on an isolated square lattice at $K=0.5$ \cite{szolnoki_pre09c}.}
\label{r2c}
\end{figure}

Quantifying accurately the impact of different values of $\alpha$ are results presented in Fig.~\ref{r2c}, where critical $r2_c$ values, along with the thresholds for cooperator and defector dominance as obtained on a single square lattice, are depicted. As already indicated in Fig.~\ref{alpha}, the largest impact on the evolution of public cooperation is obtained when the bias in the utility function is the strongest (in the vicinity of $\alpha = 0$). Here $r2_c$ approaches $r1$, indicating that the population experiences a discontinuous transition from a pure D to a pure C phase. Note that the evolution on network A becomes totally random at $\alpha=0$, because changes in the player's strategy and its utility become completely independent. In the opposite limit, when $\alpha = 0.5$, the evolution on both networks becomes statistically identical, and moreover, is almost the same as reported previously for a single square lattice. Consequently, $r2_c(\alpha=0.5) \approx r2$, indicating that to connect two graphs by means of a symmetric utility function will not result in a significant change of the behavior that is principally determined by the topology of a single graph.

\begin{figure}
\centerline{\scalebox{0.5}[0.5]{\includegraphics{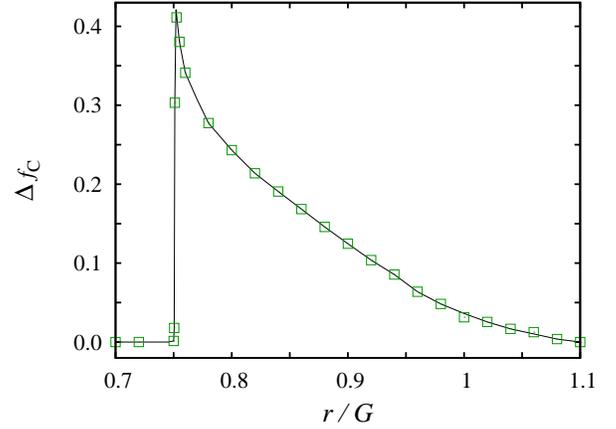}}}
\caption{Difference between the aggregated level of cooperation on both interdependent networks and the level of cooperation as obtained for an isolated square lattice $\Delta f\rm{_C}$ in dependence on the normalized synergy factor $r/G$, as obtained for $\alpha=0.01$.}
\label{delta}
\end{figure}

Since it would be possible that the biased consideration of payoffs in the utility function promotes the evolution of cooperation only on network A, while at the same time potentially having negative consequences for public cooperation on network B, it is also instructive to examine the aggregate improvement in the evolution of public cooperation. Especially so to eliminate possible doubts related to whether the interdependence truly promotes cooperative behavior, or maybe it rather just rearranges the strategies, while in fact the overall level of cooperation on both networks is determined exclusively by the value of $r$ as on a single (isolated) network. For this purpose, we plot in Fig.~\ref{delta} the difference between the overall level of cooperation and the level of cooperation as obtained for a single square lattice. As the figure shows, the promotion of cooperation is indeed a real consequence of the interdependence by means of the biased utility function. The averaged level of cooperation on both interdependent networks exceeds the level observed on a single square lattice across the whole span of $r$ values where a mixed C+D phase is possible. Note that the impact of biased utility becomes negligible on network B (A) if the population on the network A (B) arrives at an ordered (full D or full C) state. In that case the evolution on network B (A) becomes similar to that on a single network because all the players will gain the same additional payoff from the corresponding players on network A (B). In this way the feedback between the strategy change and the local success of a player is recovered. However, if staying in the mixed strategy region, the support is the strongest when the conditions for the survival of cooperators are worst, \textit{i.e.} when the synergy factor of collaborative efforts is small.

\begin{figure}
\centerline{\scalebox{0.52}[0.52]{\includegraphics{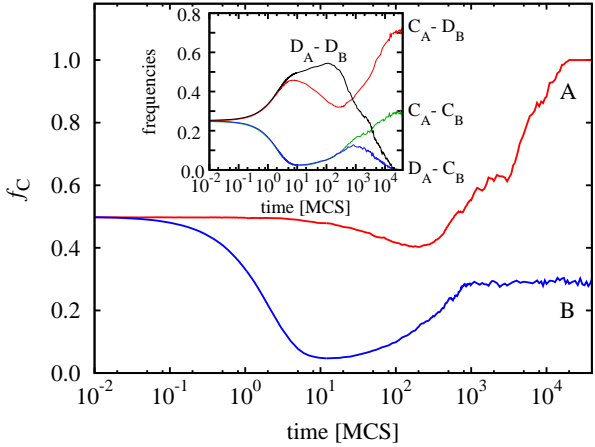}}}
\caption{Simultaneous time evolution of $f\rm{_C}$ on networks A and B (as denoted), obtained for $\alpha=0.01$ and $r/G=0.76$ if starting from a random initial state. Inset shows the evolution of frequencies for strategy pairs (as denoted) of corresponding players on the two networks.}
\label{time}
\end{figure}

To understand the origin of the reported promotion of cooperation we refer back to results presented in Fig.~\ref{snaps}, in particular to the large difference in the characteristic time scales related to the pace of evolution on networks A and B, which we also quantify more accurately in Fig.~\ref{time}. Starting from a random initial state, the fraction of cooperators starts decaying first. This is a well-known consequence of the random distribution of strategies, which is beneficial for defectors since they can easily exploit the vicinity of cooperators and hence spread efficiently \cite{szolnoki_epjb09, santos_jeb06}. The invasion of defectors on network A, however, is very much retarded. There the minimum of $f\rm{_C}$ is reached an order of magnitude later, but even more importantly, the minimal fraction of cooperators reached is much higher than on network B (see Fig.~\ref{time}). The bridle of the aggressive invasion of defectors is a straightforward consequence of the biased utility function, which suppresses the feedback between the strategy update and the possible payoff enlargement of a player. More precisely, strategy invasions on network A are predominantly dictated by the payoffs of the corresponding players on network B. Consequently, a defector on network A, who might take advantage from the vicinity of cooperators cannot invade efficiently, because the corresponding distribution of strategies in the same area on network B may be very different. On the other hand, on network B, where the players are focused predominantly on maximizing their own payoffs, the feedback between the dynamics of evolution and the utility function that drives this evolution remains almost completely intact. The stationary density of cooperators as a function of $r$ on network B is therefore very similar to the one reported for the traditional single network case \cite{szolnoki_pre09c}, especially for low values of $\alpha$. Surviving cooperators who manage to prevail against the initial invasions of defectors organize themselves into compact domains, thereby obtaining the support (spatial reciprocity) needed to spread in the sea of defectors. The significantly different time evolutions of $f\rm{_C}$ on the two networks are also conspicuous at this stage of the game. While the stationary mixed C+D phase on network B is reached after $\sim 10^3$ MCS, the evolution on network A is not just significantly slower (lasts ten times longer), but it is also more beneficial for cooperation. This can be further corroborated by the evolution of possible strategy pairs of corresponding players in the two networks during the microscopic organization (inset of Fig.~\ref{time}). Results indicate that cooperation start spreading only when defectors run out of cooperators to exploit, \textit{i.e.} when the number of $\rm{D_A-D_B}$ pairs reaches the maximum value. Thereafter, both $\rm{C_A-D_B}$ and $\rm{C_A-C_B}$ pairs start spreading simultaneously. It is crucial to note that cooperators on network B cannot survive if their partners (corresponding players) on network A are defectors ($\rm{D_A-C_B}$ falls). Evidently, although the slowing down of evolution by suppressed feedback is a strategy-neutral intervention into the dynamics, it still has very different consequences for the success (spreading) of the two competing strategies. Similar features were earlier observed when the strategy teaching \cite{szolnoki_pre09} or strategy learning capacities of players \cite{szolnoki_pre10b} were considered as being time-dependent. More generally, present results support the comprehensively accepted assumption that the different time scales in microscopic dynamics may relevantly influence the evolution of cooperation in complex systems \cite{roca_prl06, pacheco_jtb06, pacheco_prl06, szolnoki_epjb09, rong_pre10, iniguez_pre11, liu_rr_pa10}. We conclude that to consider the more realistic interdependent networks offers a new phenomenon when spontaneous separation of time scales emerges exclusively due to the interdependence between the two networks as defined by the biased utility function.

\section{Summary}

In sum, we have shown that the study of evolution of public cooperation on interdependent networks can provide new insights as to why selfish and unrelated individuals venture into collaborative efforts in sizable groups. We have exploited the concept of utility functions to create an interdependence between the two networks, revealing that biased considerations of payoffs can lead to the spontaneous separation of characteristic time scales of evolution by means of a suppressed feedback of microscopic dynamics that governs the strategy changes. In so doing, the invasion of defectors on the network where the pace of evolution is slowed down is obstructed by the fact that an immediate presence of cooperators on one network is not necessarily linked to a higher utility function on the other network. Since, however, the clustering of cooperators into large groups is an inherently slower evolutionary process than the aggressive invasion of individual defectors, and thus the spreading of cooperative behavior is not negatively influenced by the interdependence, the successful evolution of public cooperation is effectively enhanced. The presented results help us to understand why defection is not so successful if the utility to be maximized is determined not just based on local, \textit{i.e.} the nearest neighbors, but also on global, \textit{i.e.} players that are situated in another network, sources. Although it is in general expected that the local structure will promote the evolution of cooperation by means of spatial reciprocity, sacrificing some fraction of this effect on the expense of preventing defectors to invade effectively may yield a net advance for cooperators. The biased utility function introduced here captures succinctly such a scenario. Although our model is too simple to be directly applicable to a concrete situation, it is nevertheless capable to capture the essence of an interesting everyday example. This has to do with the fact that humans often rely on ``governmental'' sources, and that thus the local and global interests are not necessarily strongly correlated. Withholding taxes or exploiting social security (assuming within reason) will hardly affect our interactions with others within our immediate neighborhood. Hence, we may be tempted to cheat on sources that appear distant or not directly related to our primary activities. Creating a reliable and robust interdependence between the two networks, as we demonstrate in this letter, may then provide the necessary leverage to elevate the awareness of the bigger picture and thus raise the level of cooperation in the society.

\acknowledgments

Authors acknowledge support from the Hungarian National Research Fund (grant K-73449) and the Slovenian Research Agency (grant J1-4055).

\end{document}